\documentclass[pdflatex,sn-mathphys-num,iicol]{sn-jnl}
\usepackage{graphicx}%
\usepackage{multirow}%
\usepackage{amsmath,amssymb,amsfonts}%
\usepackage{amsthm}%
\usepackage{mathrsfs}%
\usepackage[title]{appendix}%
\usepackage{xcolor}%
\usepackage{textcomp}%
\usepackage{manyfoot}%
\usepackage{booktabs}%
\usepackage{algorithm}%
\usepackage{algorithmicx}%
\usepackage{algpseudocode}%
\usepackage{listings}%
\theoremstyle{thmstyleone}%
\theoremstyle{thmstyletwo}%

\theoremstyle{thmstylethree}%

\begin{document}

\title[Entanglement in VQC for Breast Cancer Diagnosis]{Quantum Entanglement in Variational Quantum Classification for Breast Cancer Diagnosis}

\author*[1]{\fnm{Zineb} \sur{Hazmoun}}\email{zineb.hazmoun-etu@etu.univh2c.ma}

\author[1]{\fnm{Zoubida} \sur{Sakhi}}\email{zb.sakhi@gmail.com}

\author[1]{\fnm{Mohamed} \sur{Bennai}}\email{mohamed.bennai@univh2c.ma}

\affil[1]{%
    \orgdiv{Quantum Physics and Magnetism Team, LPMC},
    \orgname{Faculty of Sciences Ben M'sick, Hassan II University of Casablanca},
    \orgaddress{\city{Casablanca}, \country{Morocco}}%
}

\abstract{This study looks at how the entangling structure of a variational quantum classifier (VQC) relates to its performance in breast cancer diagnosis, using the Wisconsin Diagnostic Breast Cancer (WDBC) dataset. We tested three three-qubit configurations that share the same EfficientSU2 ansatz, COBYLA optimizer, and stratified five-fold cross-validation, but differ in their entangling structure: a single-repetition Ising-type coupling (A), a two-repetition linear Ising-type coupling (B), and a fully connected Heisenberg-type coupling (C). Entanglement was measured with the von Neumann entropy and the Wootters concurrence. From A to C, the mean entropy rose from 0.40 to 0.71, accuracy rose from 92.98\% to 93.68\%, and F1-score rose from 90.32\% to 91.28\%. Configuration C was also the most stable across folds. However, the fold-level correlations between entanglement and performance were weak and not significant. Richer topologies also increase circuit expressivity, so the two effects cannot be fully separated. These results show an association, not a proven causal link, between entangling structure and VQC performance.}


\maketitle

\section{Introduction}
\label{sec:introduction}

Quantum computing started with Richard Feynman, who pointed out that quantum systems are hard to simulate efficiently on classical computers \cite{feynman1982}. As quantum technologies developed, these ideas were applied to machine learning, which led to quantum machine learning (QML). QML combines quantum computing with machine learning to study new ways of representing and processing data \cite{biamonte2017quantum}.

QML relies on superposition, entanglement, and interference. These properties are controlled with quantum circuits, which are made of quantum gates arranged in a given architecture. The gate structure, qubit connectivity, and depth of a circuit decide which quantum states it can produce. The architecture therefore affects not only how information is processed but also the correlations in the resulting quantum state.

Variational quantum classifiers (VQCs) are a common approach to supervised classification in QML. A VQC pairs a parameterized quantum circuit with a classical optimizer that tunes the circuit parameters for a given task. Variational quantum algorithms in general depend on this iterative loop, so the choice of ansatz is important \cite{cerezo2021variational}. QML can also be seen as a feature-space method, where classical data are encoded into quantum states and analyzed in a high-dimensional Hilbert space \cite{schuld2019quantum}. This makes the VQC a natural model for studying how changes in circuit architecture relate to quantum-state properties and to classification results.

Healthcare is a natural field for QML because medical datasets often contain complex relations among clinical, biological, and imaging variables. Breast cancer classification, where the task is to separate benign from malignant cases, is a useful test case. We use the WDBC dataset, which has 569 samples described by 30 numerical features computed from digitized images of fine-needle aspiration (FNA) samples. Its structured, low-dimensional form suits models with few qubits and allows the circuit architecture to be changed in a controlled way~\cite{memon2019}.

Entanglement is a central property of multi-qubit states, and its role in QML has drawn growing attention. Quantum feature maps, which embed classical data into quantum states, have been widely studied in relation to high-dimensional Hilbert spaces~\cite{schuld2015introduction,Hubregtsen2021}. The expressive power and trainability of parameterized circuits also depend on circuit structure and ansatz design~\cite{cerezo2021variational,holmes2022connecting}.

The link between entanglement and learning performance is not necessarily direct. A change in circuit structure can alter the expressibility, the parameter interactions, the optimization landscape, and the quantum state all at once. Theoretical work has shown, for example, that highly expressive circuits can have harder optimization landscapes~\cite{PhysRevA.111.052403}. More entanglement or a more complex circuit should therefore not be taken as a sign of better classification.

\begin{figure*}[t]
\centering
\includegraphics[width=0.7\textwidth]{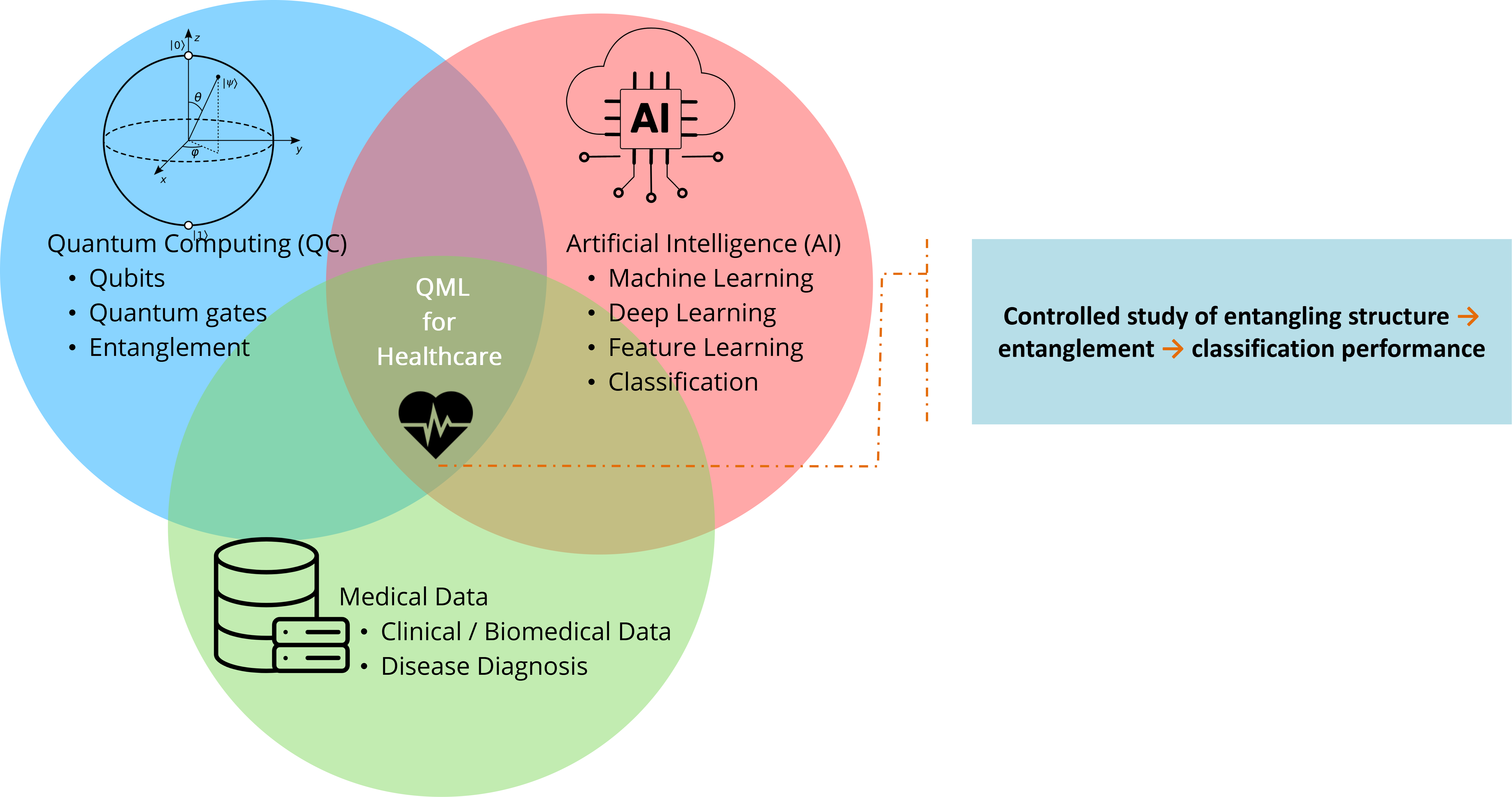}
\caption{Conceptual framework of QML for healthcare, linking quantum computing, AI, and medical data to the study of circuit structure and entanglement in medical classification.}
\label{fig:research_gap}
\end{figure*}

Earlier QML studies show that the choice of feature map and ansatz affects predictive performance. Yet comparing circuits only by their accuracy does not show how the entangling structure changes the quantum states that the model produces~\cite{el2026comparative}. This is the gap illustrated in Fig.~\ref{fig:research_gap}. Feature representations, entanglement, expressivity, and QML performance have each been studied, but few works isolate the effect of the entangling structure itself. It is still unclear whether increasing the connectivity of entangling gates changes measurable entanglement in a consistent way, and whether such changes go along with changes in classification performance when other conditions are kept similar. Since changing the connectivity affects several properties at once, a rise in entanglement together with better performance cannot prove that entanglement is the cause. A controlled design is needed to study the association between circuit-level entangling structure, measurable quantum correlations, and predictive behavior.

To address this, we evaluate three three-qubit VQC configurations with increasing entangling connectivity: a single-repetition Ising-type configuration with minimal entanglement, a two-repetition linear Ising-type configuration, and a fully connected Heisenberg-type configuration. All three use the same data representation, EfficientSU2 ansatz, COBYLA optimizer, and stratified five-fold cross-validation. Entanglement is measured with the entanglement entropy and the concurrence, and is compared with classification performance and stability. Our aim is not to show that entanglement causes better classification, but to check whether changes in entangling structure are consistently linked to changes in entanglement and predictive behavior under controlled conditions.

The rest of the paper is organized as follows. Section~\ref{sec:theory} reviews the theory of entanglement and its measures. Section~\ref{sec:qml} describes the VQC and quantum feature maps. Section~\ref{sec:methods} presents the dataset, the circuit configurations, and the experimental protocol. Section~\ref{sec:results} reports the results, including the statistical analysis. Section~\ref{sec:discussion} discusses them in light of the literature, and Section~\ref{sec:conclusion} concludes.

%%%%%%%%%%%%%%%%%%%%%%%%%%%%%%%%%%%%%%%%%%%%%%%%%%%%%%%%%%%%%
% Theoretical background
%%%%%%%%%%%%%%%%%%%%%%%%%%%%%%%%%%%%%%%%%%%%%%%%%%%%%%%%%%%%%

\section{Theoretical Background}
\label{sec:theory}

\subsection{Qubits and Multi-Qubit Hilbert Spaces}

A single qubit is a two-level quantum system. Its state is a normalized vector in the Hilbert space $\mathcal{H} = \mathbb{C}^2$,
\begin{equation}
|\psi\rangle = \alpha|0\rangle + \beta|1\rangle, \qquad |\alpha|^2+|\beta|^2=1,
\label{eq:qubit}
\end{equation}
where $\{|0\rangle,|1\rangle\}$ is the computational basis. For a register of $n$ qubits, the joint state lives in the tensor-product space
\begin{equation}
\mathcal{H}_n = \mathcal{H}^{\otimes n} = \underbrace{\mathbb{C}^2 \otimes \mathbb{C}^2 \otimes \cdots \otimes \mathbb{C}^2}_{n \text{ factors}},
\label{eq:hilbert}
\end{equation}
which has dimension $2^n$. This exponential growth of the Hilbert space with the number of qubits is what QML algorithms try to use to enrich the feature space~\cite{biamonte2017quantum}.

\subsection{Separable and Entangled States}

A pure state $|\psi\rangle \in \mathcal{H}_A \otimes \mathcal{H}_B$ of a bipartite system is called \emph{separable} (or a product state) if it can be written as
\begin{equation}
|\psi\rangle = |\phi_A\rangle \otimes |\phi_B\rangle,
\label{eq:separable}
\end{equation}
with $|\phi_A\rangle \in \mathcal{H}_A$ and $|\phi_B\rangle \in \mathcal{H}_B$. If no such factorization exists, the state is \emph{entangled}. The standard example is the Bell state
\begin{equation}
|\Phi^{+}\rangle = \frac{1}{\sqrt{2}}\big(|00\rangle + |11\rangle\big),
\label{eq:bell}
\end{equation}
which cannot be written as a product of single-qubit states. Measuring the two qubits always gives perfectly correlated outcomes, however far apart they are. This kind of correlation has no classical counterpart, and entanglement is often regarded as a key resource behind quantum computational advantage~\cite{amico2008entanglement}.

\subsection{Von Neumann Entanglement Entropy}

For a pure state $|\psi\rangle$ of a bipartite system $A\!-\!B$, the entanglement between $A$ and $B$ is measured by the von Neumann entropy of the reduced density matrix~\cite{bennett1996concentrating}. The reduced density matrix of subsystem $A$ is obtained by tracing out subsystem $B$,
\begin{equation}
\rho_A = \mathrm{Tr}_B\big(|\psi\rangle\langle\psi|\big),
\label{eq:reduced}
\end{equation}
and the entanglement entropy is
\begin{equation}
S(\rho_A) = -\mathrm{Tr}\big(\rho_A \log_2 \rho_A\big) = -\sum_i \lambda_i \log_2 \lambda_i,
\label{eq:entropy}
\end{equation}
where $\{\lambda_i\}$ are the eigenvalues of $\rho_A$ (the squared Schmidt coefficients of $|\psi\rangle$). We have $S(\rho_A)=0$ if and only if $|\psi\rangle$ is separable, and $S(\rho_A)$ reaches its maximum, $\log_2\dim\mathcal{H}_A$, for maximally entangled states such as the Bell state of Eq.~\eqref{eq:bell}. In this work, $S(\rho_A)$ is averaged over all single-qubit bipartitions of the $n$-qubit output state, which gives one entropy value per configuration~\cite{nielsen2001quantum}.

\subsection{Concurrence}

The entropy of Eq.~\eqref{eq:entropy} is defined for bipartite pure states. The \emph{concurrence} introduced by Wootters also works for mixed two-qubit states $\rho$, which makes it well suited to measuring pairwise entanglement inside a larger register. It is defined as
\begin{equation}
C(\rho) = \max\big(0,\ \lambda_1-\lambda_2-\lambda_3-\lambda_4\big),
\label{eq:concurrence}
\end{equation}
where $\lambda_1\ge\lambda_2\ge\lambda_3\ge\lambda_4$ are the square roots of the eigenvalues of the non-Hermitian matrix $\rho\tilde\rho$, and
\begin{equation}
\tilde\rho = (\sigma_y\otimes\sigma_y)\,\rho^{*}\,(\sigma_y\otimes\sigma_y),
\label{eq:rhotilde}
\end{equation}
with $\sigma_y$ the Pauli-$y$ matrix and $\rho^{*}$ the complex conjugate of $\rho$ in the computational basis. The concurrence satisfies $C(\rho)\in[0,1]$, with $C=0$ for separable states and $C=1$ for maximally entangled two-qubit states. In our experiments, the concurrence is averaged over all qubit pairs of the output state. It measures pairwise entanglement and so complements the entropy of Eq.~\eqref{eq:entropy}~\cite{wootters1998entanglement}.

\subsection{Encoding Classical Data with Controlled Entanglement}

A feature map encodes a classical feature vector $\mathbf{x}\in\mathbb{R}^d$ into a quantum state $|\Phi(\mathbf{x})\rangle = U_\Phi(\mathbf{x})|0\rangle^{\otimes n}$ through a parameterized unitary $U_\Phi$. A widely used family is the second-order Pauli-$Z$ evolution (ZZ) feature map~\cite{havlicek2019supervised}:
\begin{equation}
U_\Phi(\mathbf{x}) = \exp\!\left(i\sum_{j=1}^{n}\phi_j(\mathbf{x}) Z_j + i\sum_{j<k}\phi_{jk}(\mathbf{x}) Z_j Z_k\right)H^{\otimes n},
\label{eq:zzmap}
\end{equation}
where $H$ is the Hadamard gate, $\phi_j(\mathbf{x})=x_j$ and $\phi_{jk}(\mathbf{x})=(\pi-x_j)(\pi-x_k)$ are data-dependent phases, and the $Z_jZ_k$ terms are built from CNOT--$R_z$--CNOT blocks. The \emph{connectivity} of the $Z_jZ_k$ terms, that is, which qubit pairs are entangled and how deep the circuit is, is a free design choice. It directly affects the expressibility and the entangling capability of the circuit~\cite{sim2019expressibility}. Without any $Z_jZ_k$ term the feature map is fully separable ($S=0$, $C=0$). With nearest-neighbor terms only, it has low entanglement and linear connectivity. With all-to-all connectivity and repeated entangling layers, it has high entanglement. The three configurations studied here (A, B, C) cover this range, from a single-repetition nearest-neighbor coupling (A) to a fully connected Heisenberg-type structure (C). None of them is strictly separable, since even A keeps one nearest-neighbor entangling layer in both the feature map and the ansatz.

%%%%%%%%%%%%%%%%%%%%%%%%%%%%%%%%%%%%%%%%%%%%%%%%%%%%%%%%%%%%%
% QML framework
%%%%%%%%%%%%%%%%%%%%%%%%%%%%%%%%%%%%%%%%%%%%%%%%%%%%%%%%%%%%%

\section{Quantum Machine Learning }
\label{sec:qml}

\subsection{Variational Quantum Classifier}

A VQC combines the feature map $U_\Phi(\mathbf{x})$ of Eq.~\eqref{eq:zzmap} with a trainable variational circuit $U_W(\boldsymbol{\theta})$~\cite{farhi2018classification}. The variational circuit is made of parameterized single-qubit rotations and entangling operations, so it can adapt the encoded representation to the classification task. Here it follows the EfficientSU2 architecture, a hardware-efficient ansatz with alternating layers of single-qubit rotations and entangling blocks~\cite{kandala2017hardware}. The final state is measured with an observable $\mathcal{O}$, and the model output is

\begin{equation}
f(\mathbf{x};\boldsymbol{\theta}) =
\langle 0|^{\otimes n}
U_\Phi^\dagger(\mathbf{x})
U_W^\dagger(\boldsymbol{\theta})
\,\mathcal{O}\,
U_W(\boldsymbol{\theta})
U_\Phi(\mathbf{x})
|0\rangle^{\otimes n},
\label{eq:vqc}
\end{equation}

where $\mathcal{O}$ is the observable used to read out the class. During training, the parameters $\boldsymbol{\theta}$ are updated step by step to reduce the classification loss. This is a hybrid quantum--classical procedure: the quantum circuit gives the objective function, and a classical routine updates the parameters. We use COBYLA~\cite{powell1994direct} and keep the ansatz and the optimizer the same for all feature-map configurations. In this way, any effect of the feature-map structure and its entanglement can be studied without extra changes coming from the variational circuit or the optimization~\cite{cerezo2021variational}.

%%%%%%%%%%%%%%%%%%%%%%%%%%%%%%%%%%%%%%%%%%%%%%%%%%%%%%%%%%%%%
% Materials and Methods
%%%%%%%%%%%%%%%%%%%%%%%%%%%%%%%%%%%%%%%%%%%%%%%%%%%%%%%%%%%%%

\section{Materials and Methods}
\label{sec:methods}

\subsection{Dataset}

We use the Wisconsin Diagnostic Breast Cancer (WDBC) dataset~\cite{wolberg1990multisurface,street1993nuclear}. It contains 569 instances (357 benign and 212 malignant), each described by 30 real-valued features computed from digitized images of fine-needle aspirates of breast masses. As explained in Section~\ref{subsec:protocol}, the 30 standardized features were first filtered with Pearson correlation and then reduced by Principal Component Analysis (PCA) to three components, one for each qubit of the feature map. All preprocessing was done with scikit-learn~\cite{pedregosa2011scikit} and fitted separately inside each cross-validation fold, so that no information from the validation data leaks into training.

\subsection{Entanglement Configurations}

To study the effect of entanglement on classification, we built three Hamiltonian-inspired feature maps in Qiskit~\cite{abraham2019qiskit}. They differ in the interaction model and in the connectivity and depth of the entangling gates, while the number of qubits, the ansatz, and the optimizer stay fixed:
\begin{itemize}
\item \textbf{A -- Ising-type feature map, single repetition (minimal entanglement):} an Ising-type feature map with one repetition ($\mathrm{reps}=1$), combined with an EfficientSU2 ansatz with linear entangling connectivity and one repetition. Both parts contain nearest-neighbor $ZZ$-type entangling gates, so A is not a zero-entanglement circuit. It is the configuration with the fewest entangling layers and serves as the low-entanglement reference for B and C.
\item \textbf{B -- Ising-type feature map, two repetitions (intermediate entanglement):} the same feature map as A, repeated twice ($\mathrm{reps}=2$), with an EfficientSU2 ansatz of linear connectivity and two repetitions. This doubles the number of nearest-neighbor entangling layers relative to A.
\item \textbf{C -- Heisenberg-type feature map, two repetitions (full entanglement):} a Heisenberg-type feature map with $XX+YY+ZZ$ interaction terms, all-to-all connectivity, and two repetitions, combined with an EfficientSU2 ansatz with full entangling connectivity and two repetitions. It has the richest entangling structure of the three.
\end{itemize}

\begin{figure*}[!t]
    \centering
    \includegraphics[width=\textwidth]{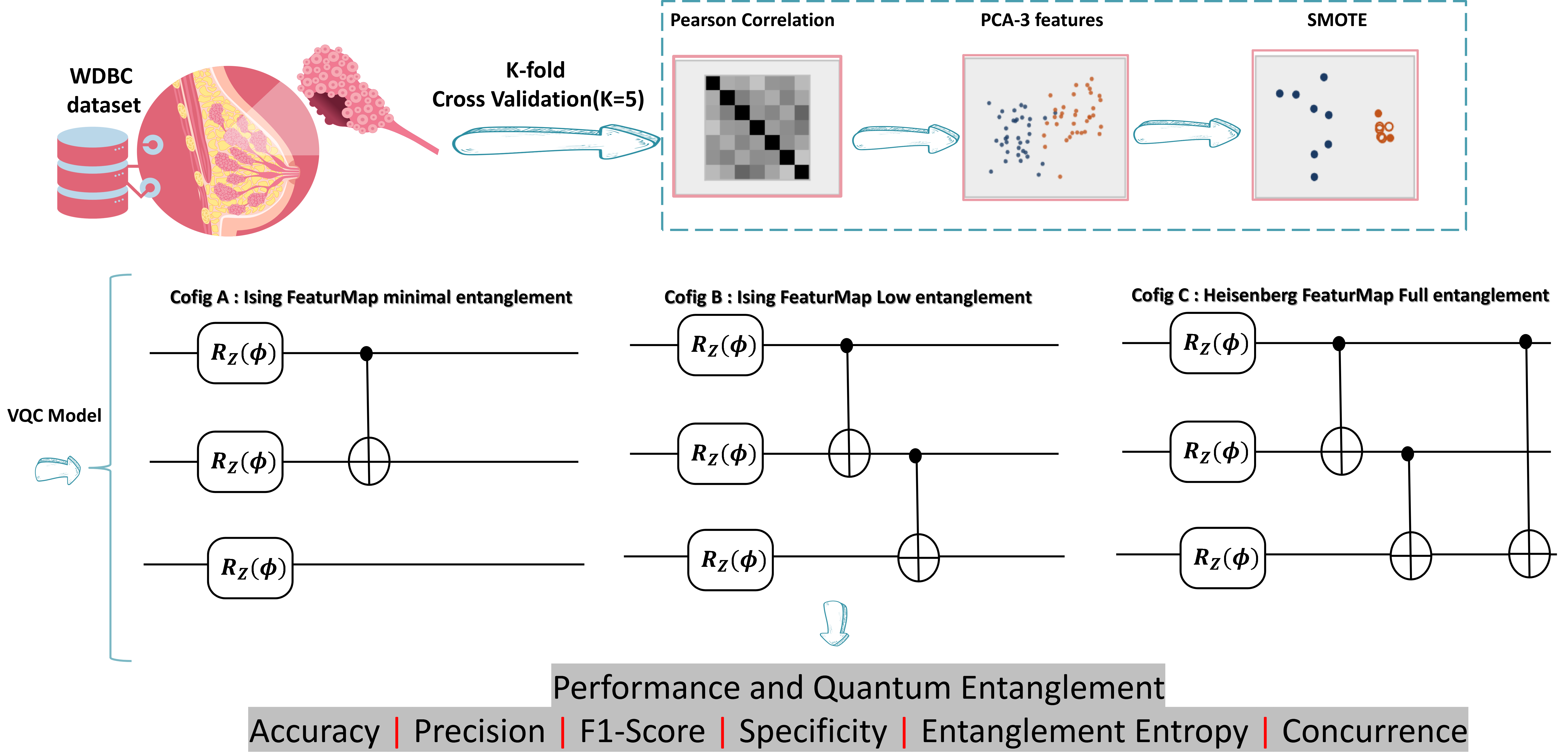}
    \caption{Overall benchmark pipeline, from preprocessing to the evaluation of configurations A, B, and C using classification and entanglement metrics.}
    \label{fig:benchmark}
\end{figure*}

For each configuration and each of the five folds, we computed the average von Neumann entanglement entropy (Eq.~\eqref{eq:entropy}, denoted EE) and the average pairwise concurrence (Eq.~\eqref{eq:concurrence}, denoted Conc). Both were obtained with Qiskit's statevector simulator on a sample of test points, using the state produced by the \emph{full trained circuit}, that is, the feature map followed by the optimized ansatz with its trained parameters. The reported entanglement therefore reflects the whole VQC pipeline and not the feature map alone.

\subsection{Training Protocol}
\label{subsec:protocol}

A preprocessing and feature-selection pipeline was applied before classification. Pearson correlation analysis kept the most relevant features and removed redundant ones. PCA then reduced the selected features to a compact form suitable for quantum encoding. To handle class imbalance, the Synthetic Minority Over-sampling Technique (SMOTE)~\cite{chawla2002smote} was applied only to the training data of each fold, so the validation data stay untouched.

Classification used a VQC with the EfficientSU2 ansatz and the COBYLA optimizer, with a fixed budget of 150 iterations, inside a stratified five-fold cross-validation ($k=5$). In each fold, all preprocessing steps, including SMOTE, were fitted on the training subset only, and the trained model was evaluated on the held-out subset. Performance was measured by accuracy, precision, recall, F1-score, and specificity. In addition, the quantum states were characterized by EE and Conc, which give two complementary views of the entanglement produced by each configuration. Computation time was not profiled here and is left for future benchmarking, which could also include noise-aware optimizers for NISQ-era training~\cite{larson2025novel}.

Results for each configuration (A, B, C) are given as mean $\pm$ standard deviation over the five folds. Fig.~\ref{fig:benchmark} summarizes the whole procedure. All experiments ran on Qiskit's noiseless statevector simulator, with one fixed random seed for data splitting, SMOTE, and parameter initialization. There was no hardware noise, no averaging over seeds, and no independent dataset for replication. These points are discussed as limitations in Section~\ref{subsec:limitations}.

%%%%%%%%%%%%%%%%%%%%%%%%%%%%%%%%%%%%%%%%%%%%%%%%%%%%%%%%%%%%%
% Results
%%%%%%%%%%%%%%%%%%%%%%%%%%%%%%%%%%%%%%%%%%%%%%%%%%%%%%%%%%%%%

\section{Results}
\label{sec:results}

\subsection{Entanglement Generated by the Three Configurations}
\label{subsec:entanglement}

\begin{figure*}[!t]
\centering
\includegraphics[width=0.7\textwidth]{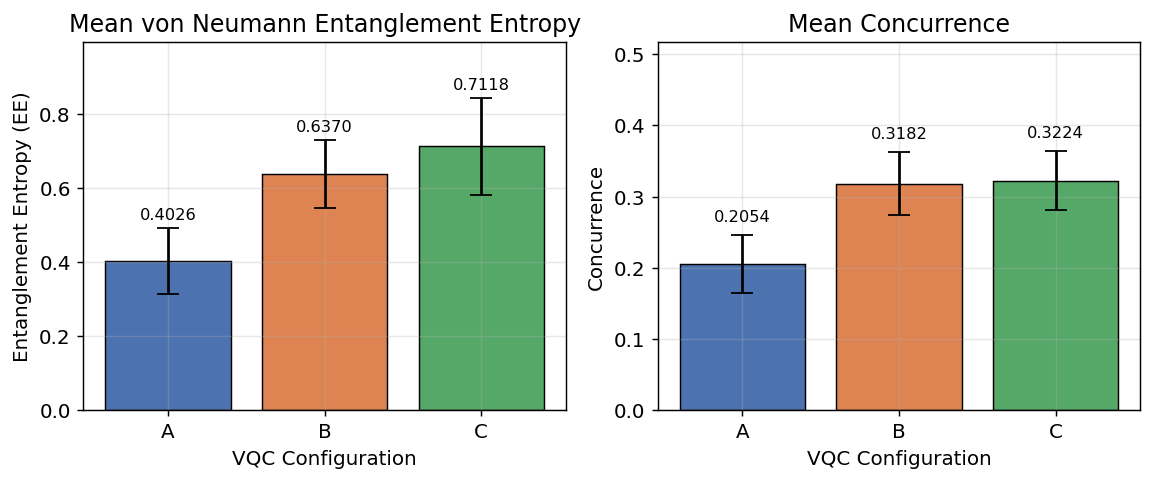}
\caption{Mean von Neumann entropy (left) and mean concurrence (right) across the five folds, for configurations A, B, and C.}
\label{fig:concurrence}
\end{figure*}

Fig.~\ref{fig:concurrence} shows that both entanglement measures increase from configuration A to B to C. The mean von Neumann entropy goes from $0.4026$ for A to $0.6370$ for B and $0.7118$ for C. The mean concurrence goes from $0.2054$ to $0.3182$ and $0.3224$. Both measures keep the order $A < B < C$, but they behave differently at the high end: the entropy still rises from B to C, while the concurrence hardly changes. The two measures thus give complementary views of the correlations.

The non-zero values for A follow from its circuit. Its feature map and its ansatz both contain nearest-neighbor $ZZ$ entangling gates, and the entanglement is measured on the full trained circuit. A is therefore not a separable, zero-entanglement control. Its values mark the lower end of the range of entangling connectivity that was explored, and the comparison focuses on the order of the configurations.

\subsection{Classification Performance }
\label{subsec:classification}

\begin{figure*}[t]
\centering
\includegraphics[width=\textwidth]{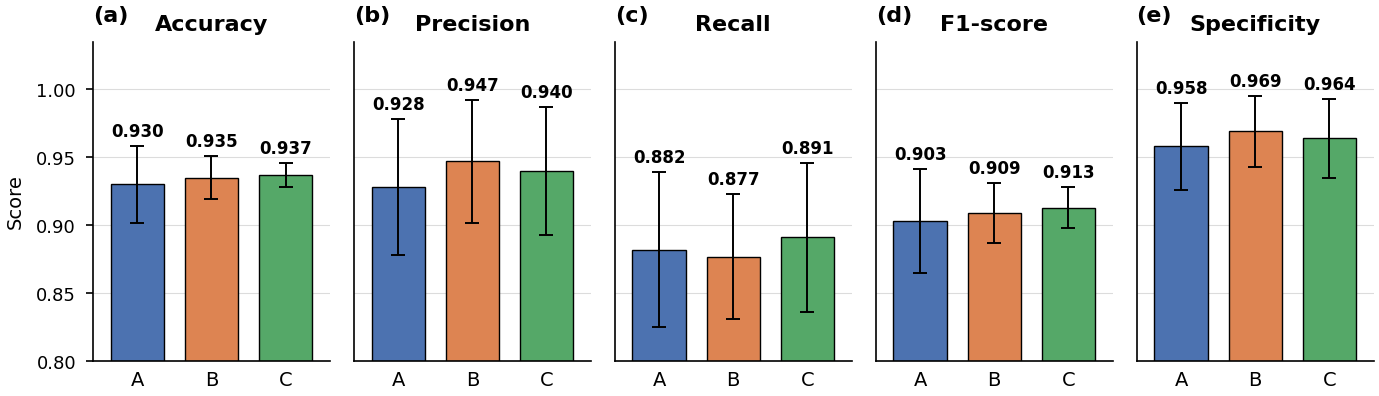}
\caption{Classification performance of the VQC for configurations A, B, and C (mean $\pm$ std. over five folds). The y-axis is zoomed to $[0.80,1.00]$ to highlight the small inter-configuration differences.}
\label{fig:performance}
\end{figure*}

\begin{figure}[!t]
\centering
\includegraphics[width=1.01\columnwidth]{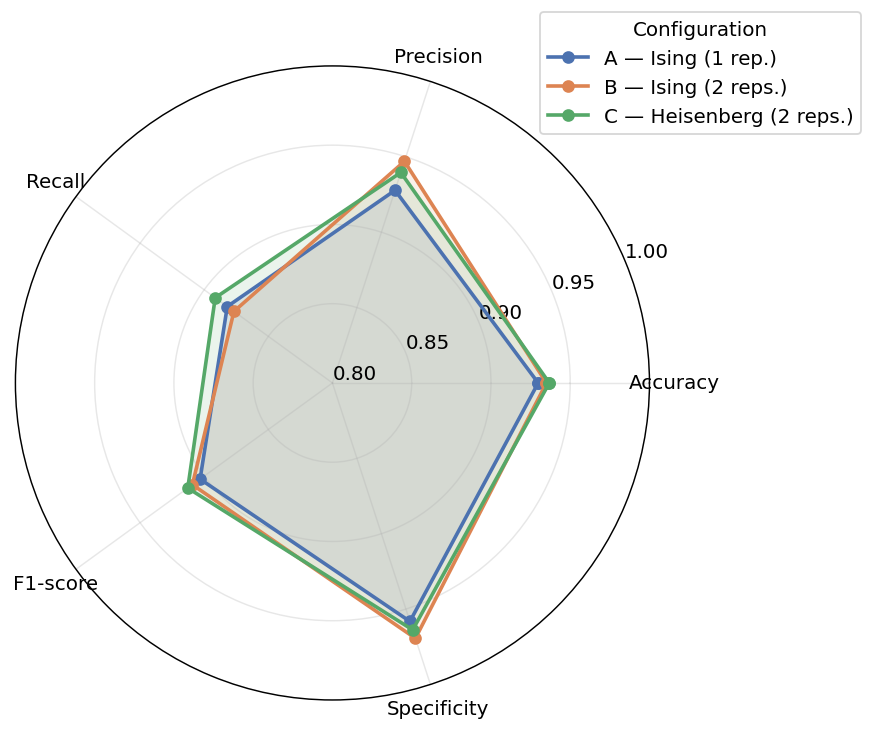}
\caption{Radar-plot comparison of the classification performance of configurations A, B, and C.}
\label{fig:radar}
\end{figure}

Table~\ref{tab:fold_results} shows that all three configurations classify well in every fold, with accuracy reaching about 0.97 (97\%) in the best case. According to Table~\ref{tab:summary_results}, the mean accuracy grows from A to B and C, with values of $0.9298 \pm 0.0278$, $0.9350 \pm 0.0160$, and $0.9368 \pm 0.0093$. The mean F1-score follows the same trend: $0.9032 \pm 0.0385$ for A, $0.9092 \pm 0.0223$ for B, and $0.9128 \pm 0.0152$ for C. Configuration C has the highest mean recall ($0.8914$), while B has the highest mean precision ($0.9470$) and specificity ($0.9694$). Fig.~\ref{fig:performance} compares the five metrics, and Fig.~\ref{fig:radar} gives a radar view of the same comparison.


% ============================================================
% TABLE: FOLD-LEVEL RESULTS
% ==========================================================
\begin{table*}[!t]
\centering
\caption{Fold-level classification performance and entanglement
metrics of the three VQC configurations evaluated on the WDBC dataset
using five-fold stratified cross-validation.}
\label{tab:fold_results}
\small
\resizebox{\textwidth}{!}{%
\begin{tabular}{@{}llccccccc@{}}
\toprule
\textbf{Configuration} &
\textbf{Fold} &
\textbf{Accuracy} &
\textbf{Precision} &
\textbf{Recall} &
\textbf{F1-score} &
\textbf{Specificity} &
\textbf{EE} &
\textbf{Concurrence} \\
\midrule

\multirow{5}{*}{A}
& 1 & 0.974 & 0.976 & 0.953 & 0.965 & 0.986 & 0.478 & 0.265 \\
& 2 & 0.912 & 0.946 & 0.814 & 0.875 & 0.972 & 0.466 & 0.178 \\
& 3 & 0.921 & 0.946 & 0.833 & 0.886 & 0.972 & 0.434 & 0.232 \\
& 4 & 0.904 & 0.844 & 0.905 & 0.874 & 0.903 & 0.377 & 0.174 \\
& 5 & 0.938 & 0.927 & 0.905 & 0.916 & 0.958 & 0.258 & 0.178 \\

\midrule

\multirow{5}{*}{B}
& 1 & 0.956 & 0.932 & 0.953 & 0.943 & 0.958 & 0.783 & 0.342 \\
& 2 & 0.930 & 0.949 & 0.860 & 0.902 & 0.972 & 0.540 & 0.303 \\
& 3 & 0.939 & 1.000 & 0.833 & 0.909 & 1.000 & 0.614 & 0.251 \\
& 4 & 0.912 & 0.881 & 0.881 & 0.881 & 0.931 & 0.596 & 0.327 \\
& 5 & 0.938 & 0.973 & 0.857 & 0.911 & 0.986 & 0.652 & 0.368 \\

\midrule

\multirow{5}{*}{C}
& 1 & 0.947 & 0.911 & 0.953 & 0.932 & 0.944 & 0.491 & 0.283 \\
& 2 & 0.930 & 0.927 & 0.884 & 0.905 & 0.958 & 0.703 & 0.315 \\
& 3 & 0.930 & 1.000 & 0.810 & 0.895 & 1.000 & 0.824 & 0.286 \\
& 4 & 0.930 & 0.886 & 0.929 & 0.907 & 0.931 & 0.760 & 0.348 \\
& 5 & 0.947 & 0.974 & 0.881 & 0.925 & 0.986 & 0.781 & 0.380 \\

\botrule
\end{tabular}%
}

\vspace{1mm}

\begin{flushleft}
\footnotesize
\textit{EE}: entanglement entropy. Values are reported for each
validation fold. Higher values of the classification metrics indicate
better predictive performance, whereas EE and concurrence quantify
different aspects of quantum correlations generated by the VQC.
\end{flushleft}
\end{table*}

\begin{table*}[!t]
\centering
\caption{Mean $\pm$ standard deviation of classification performance
and entanglement metrics across the five stratified cross-validation
folds on the WDBC dataset.}
\label{tab:summary_results}

\resizebox{\textwidth}{!}{%
\begin{tabular}{@{}lccccccc@{}}
\toprule
\textbf{Configuration} &
\textbf{Accuracy} &
\textbf{Precision} &
\textbf{Recall} &
\textbf{F1-score} &
\textbf{Specificity} &
\textbf{EE} &
\textbf{Concurrence} \\
\midrule

A &
$0.9298 \pm 0.0278$ &
$0.9278 \pm 0.0500$ &
$0.8820 \pm 0.0573$ &
$0.9032 \pm 0.0385$ &
$0.9582 \pm 0.0324$ &
$0.4026 \pm 0.0898$ &
$0.2054 \pm 0.0411$ \\

B &
$0.9350 \pm 0.0160$ &
$0.9470 \pm 0.0449$ &
$0.8768 \pm 0.0459$ &
$0.9092 \pm 0.0223$ &
$0.9694 \pm 0.0266$ &
$0.6370 \pm 0.0910$ &
$0.3182 \pm 0.0444$ \\

C &
$0.9368 \pm 0.0093$ &
$0.9396 \pm 0.0466$ &
$0.8914 \pm 0.0548$ &
$0.9128 \pm 0.0152$ &
$0.9638 \pm 0.0287$ &
$0.7118 \pm 0.1309$ &
$0.3224 \pm 0.0415$ \\

\botrule
\end{tabular}%
}

\vspace{1mm}

\begin{minipage}{\textwidth}
\footnotesize
\textit{EE}: entanglement entropy. Results are reported as mean
$\pm$ standard deviation over five stratified folds. The classification
metrics quantify predictive performance, while EE and concurrence
characterize the quantum-correlation properties of the corresponding
VQC states.
\end{minipage}

\end{table*}

\begin{figure*}[!t]
\centering
\includegraphics[width=\textwidth]{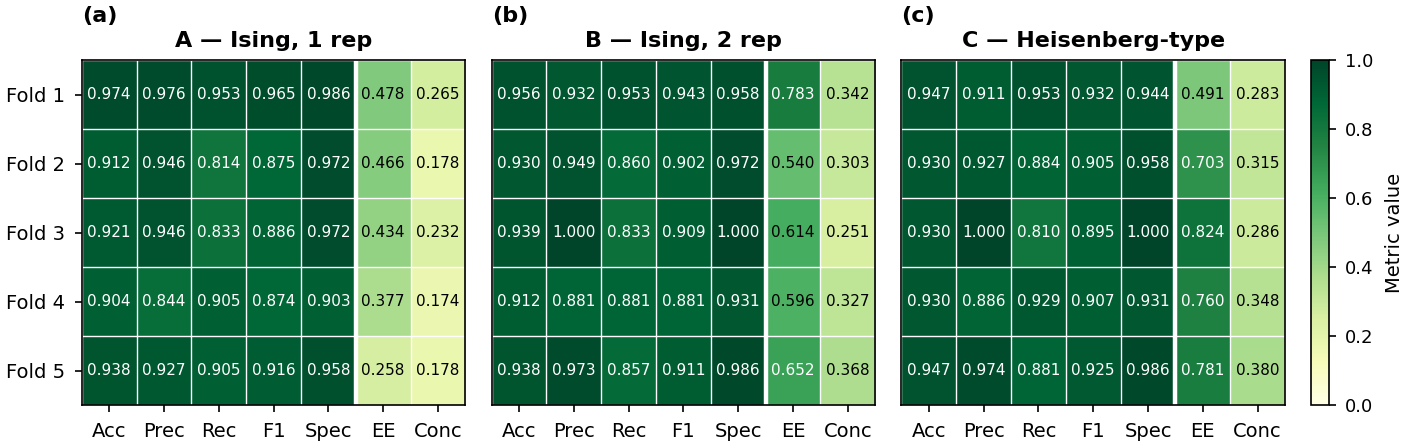}
\caption{Fold-by-fold heatmap of the classification and entanglement metrics of Table~\ref{tab:fold_results}, summarizing fold-to-fold variability for the three configurations.}
\label{fig:heatmap}
\end{figure*}

\subsection{Statistical Association Between Entanglement and Classification Performance}
\label{subsec:stats}

The complete fold-level results are given in Table~\ref{tab:fold_results} and shown fold by fold in Fig.~\ref{fig:heatmap}. Their means and standard deviations are in Table~\ref{tab:summary_results}. As the entangling structure becomes richer, both the entanglement measures and the average classification performance increase. Configuration C has the highest mean entropy ($0.7118$) and concurrence ($0.3224$), and it also has the highest mean accuracy (93.68\%) and F1-score ($0.9128$). The differences between configurations are small, though. The largest single-fold accuracy is about 97\%, but the five-fold means are the main basis for comparing A, B, and C, because they summarize all validation folds.

\begin{figure*}[!t]
\centering
\includegraphics[width=0.7\textwidth]{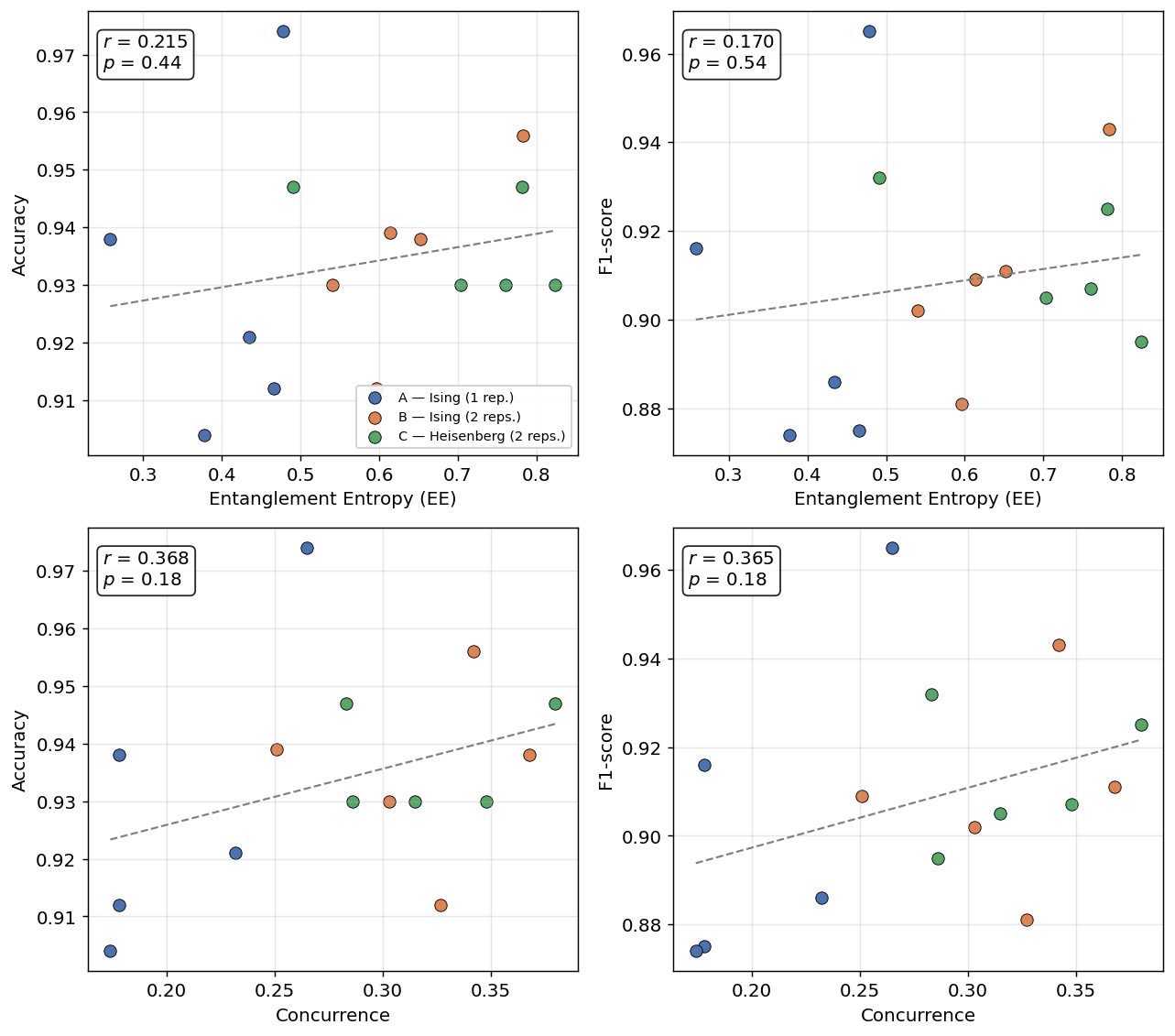}
\caption{Fold-level relationship ($n=15$) between entanglement (entropy, concurrence) and performance (accuracy, F1-score); no correlation is significant ($p>0.1$).}
\label{fig:scatter}
\end{figure*}

To measure the link between entanglement and performance, we computed Pearson correlation coefficients on the fifteen fold-level observations (three configurations, five folds each), as shown in Fig.~\ref{fig:scatter}. Between EE and accuracy, the correlation was $r=0.215$ ($p=0.44$). Between concurrence and accuracy, it was $r=0.368$ ($p=0.18$). For the F1-score, the correlations were $r=0.170$ ($p=0.54$) with EE and $r=0.365$ ($p=0.18$) with concurrence. All four coefficients are positive, but none is statistically significant.

% ============================================================
% DISCUSSION
% ============================================================

\section{Discussion}
\label{sec:discussion}

\subsection{Impact of Entangling Circuit Structure }

Across the three configurations, a richer entangling structure goes together with higher von Neumann entropy and higher concurrence. Mean classification performance follows the same order. Accuracy rises from 92.98\% (A) to 93.50\% (B) and 93.68\% (C), and F1-score rises from 0.9032 to 0.9092 and 0.9128 (Table~\ref{tab:summary_results}, Fig.~\ref{fig:performance}). Configuration C also has the lowest standard deviation over the five folds, so its results are the most consistent in this experiment.

This agrees with the role of a feature map, which turns classical inputs into quantum states. The structure of the encoding circuit changes how the data are placed in the Hilbert space and what correlations the representation contains~\cite{schuld2019quantum,havlicek2019supervised}. Entangling gates produce non-separable representations, in which information stored on different qubits becomes correlated. Here this appears as the rise in entropy ($0.4026$, $0.6370$, $0.7118$) and in concurrence ($0.2054$, $0.3182$, $0.3224$). Since these quantities and the classification metrics increase together, the entangling structure seems to be a relevant part of the feature representation used by the VQC.

Still, the fold-level correlations between entanglement and performance are positive but not significant with fifteen observations (Section~\ref{subsec:stats}). A missing significance does not change the order of the means, but it shows that the association is only moderate. In addition, the three circuits are feature-map designs and not an experiment where entanglement is the only variable. Configuration A has a single-repetition Ising feature map and one linear entangling layer in the ansatz. Configuration B doubles the repetitions with the same linear Ising-type connectivity. Configuration C uses a fully connected Heisenberg-type structure with more entangling depth. From A to C, connectivity, interaction type, and depth all change together, so the improvement reflects their combined effect. Our results support a positive empirical role for entanglement-rich circuits in VQC classification, but not the stronger claim that entanglement alone explains the differences. This caution matters especially for small biomedical datasets, where circuit architecture, representation geometry, and data variability all influence the outcome~\cite{abbas2021power,huang2021power}.

\subsection{Entropy--Concurrence Complementarity}

The two entanglement measures do not behave in the same way. From B to C, the von Neumann entropy grows clearly, from $0.6370$ to $0.7118$, while the concurrence changes only from $0.3182$ to $0.3224$. So the two quantities are not interchangeable descriptors of the states produced by the circuits. Using a single measure could hide changes in the structure of the correlations. Here the entropy is more sensitive to the step from the intermediate to the strongly entangled configuration. Using both gives a fuller description of the states. It also agrees with the known result that expressibility and entangling capability are distinct properties of a parameterized circuit, only loosely related~\cite{sim2019expressibility}.

\subsection{Classification Behavior and Clinical Metrics}

The three configurations show different performance profiles. Accuracy and F1-score grow from A to B and C, but the other metrics do not follow the same order. Configuration B has the highest precision and specificity, while configuration C has the highest recall and F1-score, together with the lowest fold-to-fold variability. The richer configuration thus has the best overall profile, and the intermediate one keeps an advantage on some criteria. None of these differences is statistically significant with $k=5$ folds.

This matters in medical classification, because each metric answers a different question. Specificity is the share of negative cases correctly identified, and recall is the share of positive cases correctly identified. Precision is the share of predicted positives that are correct, and F1-score combines precision and recall. Which configuration is best therefore depends on the clinical goal and not on accuracy alone~\cite{hicks2022evaluation}. For this reason, we recommend reporting several metrics together when VQC models are assessed for biomedical tasks~\cite{foody2023challenges}.

\subsection{Comparison with Classical Baseline Models}
\label{sec:baseline_comparison}

To put the VQC results in context, we evaluated three common classifiers on the same WDBC dataset~\cite{street1993nuclear} with exactly the same protocol as the quantum models. In each stratified fold, Pearson correlation filtering, standardization, and PCA were fitted on the training data only and reduced to three principal components, matching the three-qubit encoding. SMOTE was then applied to the training part of each fold only~\cite{chawla2002smote}. No information from the test fold therefore enters the preprocessing or the oversampling, and the same folds and preprocessing were used for all classical and quantum models.

The classical models are Logistic Regression (LR), a Support Vector Machine with a radial basis function kernel (SVM-RBF), and a Random Forest (RF) with 200 trees. They cover a linear model, a kernel-based nonlinear model, and a tree-based nonlinear model~\cite{cortes1995support,breiman2001random}. SVM-RBF is an especially useful reference for the VQC, because both can build nonlinear decision boundaries in a transformed feature space, although by very different mechanisms. Table~\ref{tab:baseline_comparison} gives the mean $\pm$ standard deviation of the five metrics over the five folds for all six models.

\begin{table*}[!t]
\centering
\caption{Comparison of classical baseline models and VQC configurations on
the WDBC dataset. Results are reported as mean $\pm$ standard deviation over
five stratified cross-validation folds. All models use the same fold-specific
preprocessing pipeline, including Pearson filtering, standardization, PCA to
three components, and SMOTE applied to the training fold only.}
\label{tab:baseline_comparison}
\resizebox{\textwidth}{!}{%
\begin{tabular}{lccccc}
\toprule
\textbf{Model} &
\textbf{Accuracy} &
\textbf{Precision} &
\textbf{Recall} &
\textbf{F1} &
\textbf{Specificity} \\
\midrule

LR &
$0.9245 \pm 0.0088$ &
$0.8839 \pm 0.0301$ &
$0.9199 \pm 0.0236$ &
$0.9009 \pm 0.0099$ &
$0.9272 \pm 0.0223$ \\

SVM-RBF &
$0.9244 \pm 0.0180$ &
$0.8809 \pm 0.0387$ &
$0.9246 \pm 0.0400$ &
$0.9010 \pm 0.0234$ &
$0.9245 \pm 0.0283$ \\

RF &
$0.9227 \pm 0.0170$ &
$0.9021 \pm 0.0551$ &
$0.8962 \pm 0.0621$ &
$0.8958 \pm 0.0227$ &
$0.9385 \pm 0.0389$ \\

\midrule

VQC-A &
$0.9298 \pm 0.0278$ &
$0.9278 \pm 0.0500$ &
$0.8820 \pm 0.0573$ &
$0.9032 \pm 0.0385$ &
$0.9582 \pm 0.0324$ \\

VQC-B &
$0.9350 \pm 0.0160$ &
$\mathbf{0.9470 \pm 0.0449}$ &
$0.8768 \pm 0.0459$ &
$0.9092 \pm 0.0223$ &
$\mathbf{0.9694 \pm 0.0266}$ \\

VQC-C &
$\mathbf{0.9368 \pm 0.0093}$ &
$0.9396 \pm 0.0466$ &
$0.8914 \pm 0.0548$ &
$\mathbf{0.9128 \pm 0.0152}$ &
$0.9638 \pm 0.0287$ \\

\botrule
\end{tabular}%
}
\end{table*}

The three classical models have very similar mean accuracies, from $0.9227$ (RF) to $0.9245$ (LR), and similar F1-scores, from $0.8958$ to $0.9010$. All three VQC configurations give higher mean accuracy and F1-score than any of them. Configuration C is the best of all models, with accuracy $0.9368 \pm 0.0093$ and F1-score $0.9128 \pm 0.0152$, which is about 1.2 percentage points above the best classical result. This gap is small and should be read carefully. Configuration A, with the fewest entangling layers, already reaches an accuracy of $0.9298 \pm 0.0278$ and an F1-score of $0.9032 \pm 0.0385$, slightly above the classical models. So the behavior of the VQC cannot be explained by entanglement alone. The parameterized feature transformation, the circuit depth, and the variational optimization may also play a part. The comparison shows a measurable difference between this family of VQCs and the chosen classical models, but it does not isolate entanglement as its source.

The VQC models also differ from the classical ones in specificity and recall. VQC-B has the highest mean specificity ($0.9694 \pm 0.0266$), followed by VQC-C ($0.9638 \pm 0.0287$), while the best classical value is $0.9385 \pm 0.0389$ for RF. In a diagnostic setting, specificity is the share of non-malignant cases correctly labeled negative, so it relates to the control of false positives. Recall is higher for the classical models, with $0.9246 \pm 0.0400$ for SVM-RBF and $0.9199 \pm 0.0236$ for LR, against $0.8914 \pm 0.0548$ for the best VQC (configuration C). The VQC models therefore do not improve every criterion. They give higher specificity and precision, and the classical models give higher recall. The weight of false positives against false negatives depends on the clinical use~\cite{saito2015precision}.

Because all models share the same folds, dimensionality, preprocessing, and training-fold-only SMOTE, the differences cannot easily be explained by data preparation or by the evaluation protocol. Including SVM-RBF also gives a nonlinear reference and not only a linear one. Even so, an improvement of about 1.2 percentage points is not evidence of quantum advantage. Showing such an advantage would need a much broader comparison with computational, statistical, and resource-level controls. Our purpose is narrower: to make sure the relation between circuit structure, entanglement, and performance is not judged against a weak classical reference. Overall, entanglement should be seen as one architectural factor among several that can influence VQC behavior and not as an isolated cause of better classification.

\subsection{Comparison with Recent QML Studies}

Our results can be placed within the growing literature on QML for biomedical and cancer classification. Most recent work aims at better predictive performance through hybrid quantum--classical models, quantum transfer learning, or variational circuits. Xiang \textit{et al.} studied a quantum--classical convolutional architecture for breast cancer diagnosis on the WDBC, GBSG, and SEER datasets~\cite{xiang2024quantum}. Azevedo \textit{et al.} applied quantum transfer learning to breast cancer detection and obtained competitive performance with a shallow quantum circuit~\cite{azevedo2022quantum}. For other cancers, Khan \textit{et al.} built a hybrid quantum convolutional neural network for brain-tumor classification from MRI data~\cite{khan2024brain}, and El Maouaki \textit{et al.} showed that a quantum support vector machine can outperform its classical counterpart for prostate cancer detection under a matched feature-map and kernel setting~\cite{elmaouaki2024quantum}. Chen \textit{et al.} examined the optimization landscape of variational quantum classifiers and stressed the role of trainability~\cite{sen2022variational}, and Zarei Ghobadi and Afsaneh reported competitive results for QML on several cancer types with real biological data~\cite{zarei2024potential}.

Unlike these studies, our main aim is not another accuracy benchmark. We ask how the structure of a quantum feature map relates to the quantum correlations produced by the circuit and to VQC classification behavior. With the same architecture, number of qubits, optimizer, and cross-validation protocol, only the entangling structure was changed step by step. This lets us look at classification behavior together with direct measurements of the generated correlations, and avoids treating the quantum circuit as a black box. The findings agree with earlier work showing that the feature-map structure shapes the quantum representation and the expressive properties of the model~\cite{havlicek2019supervised,schuld2021effect}, and that entangling gates create correlations between subsystems and so change the feature space available to the learner~\cite{abbas2021power}.

We also do not read higher performance as a sign of quantum advantage. The differences between configuration means are not statistically significant at five folds (Section~\ref{subsec:stats}), and configuration C changes not only the amount of entanglement but also the interaction type and the feature-map family. This view fits recent reviews, which find that evidence for a general quantum advantage in healthcare is still inconclusive. Maheshwari \textit{et al.} pointed to the large variety of datasets, algorithms, and evaluation methods in biomedical QML~\cite{mahashwari2022review}, and a review focused on breast cancer reached a similar conclusion~\cite{kaveh2025investigating}. Durant \textit{et al.} noted that comparisons of QML and classical
machine-learning approaches in biomedical applications remain
challenging, as reported performance depends on the datasets and
experimental settings considered~\cite{durant2024primer}. Gupta \textit{et al.} concluded that the available evidence does not yet show a consistent advantage of QML over classical methods under realistic healthcare conditions~\cite{gupta2025systematic}. Against this background, our contribution is to report accuracy, entanglement entropy, concurrence, and fold-to-fold variability together under one controlled VQC protocol, and to test the link between them beyond configuration-level averages.

\subsection{Limitations}
\label{subsec:limitations}

Several limitations should be kept in mind. First, all experiments used Qiskit's noiseless statevector simulator, with no hardware noise and no noise-aware training, so the results may not carry over to near-term quantum hardware. Second, one fixed random seed controlled data splitting, SMOTE, and parameter initialization, and results were not averaged over seeds. The reported means and standard deviations thus reflect variation across folds only, not across initializations. Third, only one dataset (WDBC) and one three-qubit encoding were used, so generalization to other datasets, qubit counts, or diseases is untested. Fourth, the correlation analysis uses $n=15$ observations from three configurations that share the same five folds. These observations are not fully independent, and no correction for multiple comparisons was made across the four correlations, so the $p$-values are exploratory and not confirmatory. Finally, entanglement was measured on the full trained circuit, and the three configurations differ in both the number of entangling layers and the total depth. Entanglement and circuit expressivity are therefore confounded, and the design does not isolate entanglement as an independent variable.

These limits also fit what is known about variational quantum algorithms in general. They are designed for near-term devices~\cite{preskill2018quantum}, yet remain sensitive to circuit architecture, optimization, trainability, and noise~\cite{cerezo2021variational,mcclean2018barren,cerezo2022challenges}. A more complex or more entangled circuit can be more expressive but also harder to optimize and more exposed to noise. The gain observed from A to C should thus be seen as a modest and statistically weak trend. It does not show that adding entanglement without limit would keep improving performance, or that the trend would hold on another dataset or with another random seed.

% ============================================================
% CONCLUSION
% ============================================================

\section{Conclusion}
\label{sec:conclusion}

This study examined how circuit-level entangling structure, quantum correlations, and classification behavior are related in a three-qubit variational quantum classifier for breast cancer diagnosis on the WDBC dataset. Three VQC configurations were compared with the same EfficientSU2 ansatz, COBYLA optimizer, three-qubit representation, and stratified five-fold cross-validation. They differ in their entangling structure: a single-repetition Ising-type configuration, a two-repetition linear Ising-type configuration, and a fully connected Heisenberg-type configuration.

The measured quantum correlations grew as the entangling structure became richer. The mean von Neumann entropy went from 0.4026 (A) to 0.6370 (B) and 0.7118 (C), and the concurrence went from 0.2054 to 0.3182 and 0.3224. Classification performance rose modestly at the same time, with accuracy going from 92.98\% to 93.50\% and 93.68\%, and F1-score from 90.32\% to 90.92\% and 91.28\%. Configuration C had the highest mean accuracy, recall, and F1-score, while configuration B had the highest precision and specificity.

These findings show that changes in entangling structure are associated with measurable changes in the quantum correlations of the VQC and in its predictive behavior. They do not show that entanglement alone causes the improvement. The configurations change connectivity, interaction type, and circuit depth at once, and the performance differences are not statistically significant in the five-fold analysis. The two entanglement measures also behave differently: the entropy keeps rising from B to C, while the concurrence hardly moves. Using both therefore gives a fuller picture of the states produced by the circuits.

Overall, this work offers a reproducible framework for studying entanglement as a circuit-level property of variational quantum classifiers in biomedical classification. It supports a balanced view in which entanglement, circuit architecture, quantum representation, and trainability are considered together, rather than the assumption that more entanglement always gives better classification. Future work will vary the entangling topology and the interaction strength while holding the other circuit parameters fixed, to separate their individual effects. Repeated cross-validation, independent datasets and test sets, and noise-aware experiments will help test how robust and general the observed relation is.

\backmatter

\section*{Statements and Declarations}

\bmhead{Competing Interests}
The authors declare that they have no competing interests.

\bmhead{Funding}
The authors received no specific funding for this work.

\bmhead{Data availability}
The Wisconsin Diagnostic Breast Cancer (WDBC) dataset analysed in this study is publicly available from the UCI Machine Learning Repository.

\bmhead{Ethics approval and consent to participate}
Not applicable. This study uses a publicly available, anonymised dataset.

\bibliography{references}

@article{feynman1982,
  author  = {Feynman, Richard P.},
  title   = {Simulating Physics with Computers},
  journal = {International Journal of Theoretical Physics},
  volume  = {21},
  pages   = {467--488},
  year    = {1982},
  doi     = {10.1007/BF02650179}
}

@article{wolberg1990multisurface,
  title={Multisurface method of pattern separation for medical diagnosis applied to breast cytology},
  author={Wolberg, William H. and Mangasarian, Olvi L.},
  journal={Proceedings of the National Academy of Sciences},
  volume={87}, number={23}, pages={9193--9196}, year={1990},
  doi={10.1073/pnas.87.23.9193}
}

@inproceedings{street1993nuclear,
  title={Nuclear feature extraction for breast tumor diagnosis},
  author={Street, W. Nick and Wolberg, William H. and Mangasarian, Olvi L.},
  booktitle={Proc. SPIE 1905, Biomedical Image Processing and Biomedical Visualization},
  pages={861--870}, year={1993},
  doi={10.1117/12.148698}
}

@article{kaveh2025investigating,
  title={Investigating the application of quantum machine learning in breast cancer: a systematic review},
  author={Kaveh, Shahrzad and Arezi, Elahe and Khedri, Zahra and Sohrabei, Solmaz},
  journal={Archives of Breast Cancer},
  volume={12}, number={2}, pages={130--142}, year={2025}
}

@article{memon2019,
  author  = {Memon, Muhammad Hammad and Li, Jian Ping and Haq, Amin Ul
             and Memon, Muhammad Hunain and Zhou, Wang},
  title   = {Breast Cancer Detection in the IoT Health Environment
             Using Modified Recursive Feature Selection},
  journal = {Wireless Communications and Mobile Computing},
  volume  = {2019},
  number  = {1},
  pages   = {5176705},
  year    = {2019},
  doi     = {10.1155/2019/5176705}
}

@article{preskill2018quantum,
  title={Quantum computing in the NISQ era and beyond},
  author={Preskill, John},
  journal={Quantum},
  volume={2}, pages={79}, year={2018},
  doi={10.22331/q-2018-08-06-79}
}

@book{nielsen2001quantum,
  title={Quantum Computation and Quantum Information},
  author={Nielsen, Michael A. and Chuang, Isaac L.},
  volume={2}, year={2001},
  publisher={Cambridge University Press}
}

@article{PhysRevA.111.052403,
  title = {Machine-learning insights into the entanglement-trainability correlation of parametrized quantum circuits},
  author = {Zhang, Shikun and Zhou, Yang and Qin, Zheng and Li, Rui and Du, Chunxiao and Xiao, Zhisong and Zhang, Yongyou},
  journal = {Phys. Rev. A},
  volume = {111},
  issue = {5},
  pages = {052403},
  numpages = {11},
  year = {2025},
  month = {May},
  publisher = {American Physical Society},
  doi = {10.1103/PhysRevA.111.052403},
  url = {https://link.aps.org/doi/10.1103/PhysRevA.111.052403}
}

@article{el2026comparative,
  title={Comparative performance analysis of quantum machine learning architectures for credit card fraud detection},
  author={El Alami, Mansour and Innan, Nouhaila and Shafique, Muhammad and Bennai, Mohamed},
  journal={Applied Intelligence},
  volume={56},
  number={3},
  pages={83},
  year={2026},
  publisher={Springer}
}

@article{Hubregtsen2021,
  author       = {Hubregtsen, Thomas and
                  Pichlmeier, Josef and
                  Stecher, Patrick and
                  Bertels, Koen},
  title        = {Evaluation of parameterized quantum circuits:
                  on the relation between classification accuracy,
                  expressibility, and entangling capability},
  journaltitle = {Quantum Machine Intelligence},
  year         = {2021},
  volume       = {3},
  number       = {1},
  eid          = {9},
  date         = {2021-03-11},
  issn         = {2524-4914},
  doi     
  = {10.1007/s42484-021-00038-w},
  url          = {https://doi.org/10.1007/s42484-021-00038-w}
}

@article{amico2008entanglement,
  title={Entanglement in many-body systems},
  author={Amico, Luigi and Fazio, Rosario and Osterloh, Andreas and Vedral, Vlatko},
  journal={Reviews of Modern Physics},
  volume={80}, number={2}, pages={517--576}, year={2008},
  doi={10.1103/RevModPhys.80.517}
}

@article{wootters1998entanglement,
  title={Entanglement of formation of an arbitrary state of two qubits},
  author={Wootters, William K.},
  journal={Physical Review Letters},
  volume={80}, number={10}, pages={2245--2248}, year={1998},
  doi={10.1103/PhysRevLett.80.2245}
}

@article{bennett1996concentrating,
  title={Concentrating partial entanglement by local operations},
  author={Bennett, Charles H. and Bernstein, Herbert J. and Popescu, Sandu and Schumacher, Benjamin},
  journal={Physical Review A},
  volume={53}, number={4}, pages={2046}, year={1996},
  publisher={APS}
}

@article{biamonte2017quantum,
  title={Quantum machine learning},
  author={Biamonte, Jacob and Wittek, Peter and Pancotti, Nicola and Rebentrost, Patrick and Wiebe, Nathan and Lloyd, Seth},
  journal={Nature},
  volume={549}, number={7671}, pages={195--202}, year={2017},
  doi={10.1038/nature23474}
}

@article{schuld2019quantum,
  title={Quantum machine learning in feature Hilbert spaces},
  author={Schuld, Maria and Killoran, Nathan},
  journal={Physical Review Letters},
  volume={122}, number={4}, pages={040504}, year={2019},
  doi={10.1103/PhysRevLett.122.040504}
}

@article{havlicek2019supervised,
  title={Supervised learning with quantum-enhanced feature spaces},
  author={Havl{\'\i}{\v{c}}ek, Vojt{\v{e}}ch and C{\'o}rcoles, Antonio D. and Temme, Kristan and Harrow, Aram W. and Kandala, Abhinav and Chow, Jerry M. and Gambetta, Jay M.},
  journal={Nature},
  volume={567}, number={7747}, pages={209--212}, year={2019},
  doi={10.1038/s41586-019-0980-2}
}

@article{farhi2018classification,
  title={Classification with quantum neural networks on near term processors},
  author={Farhi, Edward and Neven, Hartmut},
  journal={arXiv preprint arXiv:1802.06002}, year={2018}
}

@article{cerezo2021variational,
  title={Variational quantum algorithms},
  author={Cerezo, M. and Arrasmith, Andrew and Babbush, Ryan and Benjamin, Simon C. and Endo, Suguru and Fujii, Keisuke and McClean, Jarrod R. and Mitarai, Kosuke and Yuan, Xiao and Cincio, Lukasz and Coles, Patrick J.},
  journal={Nature Reviews Physics},
  volume={3}, number={9}, pages={625--644}, year={2021},
  doi={10.1038/s42254-021-00348-9}
}

@article{cerezo2022challenges,
  title={Challenges and opportunities in quantum machine learning},
  author={Cerezo, M. and Verdon, Guillaume and Huang, Hsin-Yuan and Cincio, Lukasz and Coles, Patrick J.},
  journal={Nature Computational Science},
  volume={2}, number={9}, pages={567--576}, year={2022},
  doi={10.1038/s43588-022-00311-3}
}

@article{kandala2017hardware,
  title={Hardware-efficient variational quantum eigensolver for small molecules and quantum magnets},
  author={Kandala, Abhinav and Mezzacapo, Antonio and Temme, Kristan and Takita, Maika and Brink, Markus and Chow, Jerry M. and Gambetta, Jay M.},
  journal={Nature},
  volume={549}, number={7671}, pages={242--246}, year={2017},
  doi={10.1038/nature23879}
}

@article{mcclean2018barren,
  title={Barren plateaus in quantum neural network training landscapes},
  author={McClean, Jarrod R. and Boixo, Sergio and Smelyanskiy, Vadim N. and Babbush, Ryan and Neven, Hartmut},
  journal={Nature Communications},
  volume={9}, number={1}, pages={4812}, year={2018},
  doi={10.1038/s41467-018-07090-4}
}

@article{abbas2021power,
  title={The power of quantum neural networks},
  author={Abbas, Amira and Sutter, David and Zoufal, Christa and Lucchi, Aur{\'e}lien and Figalli, Alessio and Woerner, Stefan},
  journal={Nature Computational Science},
  volume={1}, number={6}, pages={403--409}, year={2021},
  doi={10.1038/s43588-021-00084-1}
}

@techreport{powell1994direct,
  title={A direct search optimization method that models the objective and constraint functions by linear interpolation},
  author={Powell, Michael J. D.},
  institution={University of Cambridge, Department of Applied Mathematics and Theoretical Physics},
  year={1994},
  note={In Advances in Optimization and Numerical Analysis, Kluwer Academic, pp.\ 51--67},
  doi={10.1007/978-94-015-8330-5_4}
}

@article{larson2025novel,
  title={A novel noise-aware classical optimizer for variational quantum algorithms},
  author={Larson, Jeffrey and Menickelly, Matt and Shi, Jiahao},
  journal={INFORMS Journal on Computing},
  volume={37}, number={1}, pages={63--85}, year={2025}
}

@article{sim2019expressibility,
  author  = {Sim, Sukin and Johnson, Peter D. and Aspuru-Guzik, Al{\'a}n},
  title   = {Expressibility and Entangling Capability of Parameterized Quantum Circuits for Hybrid Quantum-Classical Algorithms},
  journal = {Advanced Quantum Technologies},
  volume  = {2}, number = {12}, pages = {1900070}, year = {2019},
  doi     = {10.1002/qute.201900070}
}

@article{huang2021power,
  author  = {Huang, Hsin-Yuan and Broughton, Michael and Cotler, Jordan and Chen, Sitan and Li, Jerry and Mohseni, Masoud and Neven, Hartmut and Babbush, Ryan and Kueng, Richard and Preskill, John and McClean, Jarrod R.},
  title   = {Exploring the Power of Entangled Data in Quantum Machine Learning},
  journal = {arXiv preprint arXiv:2107.05772}, year = {2021}
}

@misc{abraham2019qiskit,
  title={Qiskit: An open-source framework for quantum computing},
  author={Abraham, H{\'e}ctor and others},
  year={2019},
  doi={10.5281/zenodo.2562110}
}

@article{pedregosa2011scikit,
  title={Scikit-learn: Machine learning in {P}ython},
  author={Pedregosa, Fabian and Varoquaux, Ga{\"e}l and Gramfort, Alexandre and Michel, Vincent and Thirion, Bertrand and Grisel, Olivier and Blondel, Mathieu and Prettenhofer, Peter and Weiss, Ron and Dubourg, Vincent and others},
  journal={Journal of Machine Learning Research},
  volume={12}, pages={2825--2830}, year={2011}
}

@article{chawla2002smote,
  title={{SMOTE}: Synthetic minority over-sampling technique},
  author={Chawla, Nitesh V. and Bowyer, Kevin W. and Hall, Lawrence O. and Kegelmeyer, W. Philip},
  journal={Journal of Artificial Intelligence Research},
  volume={16}, pages={321--357}, year={2002}
}

@article{sen2022variational,
  title={Variational quantum classifiers through the lens of the Hessian},
  author={Sen, Pinaki and Bhatia, Amandeep Singh and Bhangu, Kamalpreet Singh and Elbeltagi, Ahmed},
  journal={Plos one},
  volume={17},
  number={1},
  pages={e0262346},
  year={2022},
  publisher={Public Library of Science San Francisco, CA USA}
}

@article{azevedo2022quantum,
  title={Quantum transfer learning for breast cancer detection},
  author={Azevedo, Vanda and Silva, Carla and Dutra, In{\^e}s},
  journal={Quantum Machine Intelligence},
  volume={4}, number={1}, pages={5}, year={2022},
  doi={10.1007/s42484-022-00062-4}
}

@article{durant2024primer,
  title={A primer for quantum computing and its applications to healthcare and biomedical research},
  author={Durant, Thomas JS and Knight, Elizabeth and Nelson, Brent and Dudgeon, Sarah and Lee, Seung J and Walliman, Dominic and Young, Hobart P and Ohno-Machado, Lucila and Schulz, Wade L},
  journal={Journal of the American Medical Informatics Association},
  volume={31},
  number={8},
  pages={1774--1784},
  year={2024},
  publisher={Oxford University Press}
}

@article{xiang2024quantum,
  title={Quantum classical hybrid convolutional neural networks for breast cancer diagnosis},
  author={Xiang, Qiuyu and Li, Dongfen and Hu, Zhikang and Yuan, Yuhang and Sun, Yuchen and Zhu, Yonghao and Fu, You and Jiang, Yangyang and Hua, Xiaoyu},
  journal={Scientific Reports},
  volume={14}, number={1}, pages={24699}, year={2024},
  doi={10.1038/s41598-024-74778-7}
}

@article{zarei2024potential,
  title={Potential of quantum machine learning for solving the real-world problem of cancer classification},
  author={Zarei Ghobadi, Mohadeseh and Afsaneh, Elaheh},
  journal={Discover Applied Sciences},
  volume={6}, pages={513}, year={2024},
  doi={10.1007/s42452-024-06220-6}
}

@article{mahashwari2022review,
  title={Quantum machine learning applications in the biomedical domain: A systematic review},
  author={Maheshwari, Danyal and Garcia-Zapirain, Begonya and Sierra-Sosa, Daniel},
  journal={IEEE Access},
  volume={10}, year={2022},
  doi={10.1109/ACCESS.2022.3195044}
}

@article{gupta2025systematic,
  title={A systematic review of quantum machine learning for digital health},
  author={Gupta, Riddhi S. and Wood, Carolyn E. and Engstrom, Teyl and Pole, Jason D. and Shrapnel, Sally},
  journal={npj Digital Medicine},
  volume={8}, pages={237}, year={2025},
  doi={10.1038/s41746-025-01597-z}
}

@article{schuld2021effect,
  title={Effect of data encoding on the expressive power of variational quantum-machine-learning models},
  author={Schuld, Maria and Sweke, Ryan and Meyer, Johannes Jakob},
  journal={Physical Review A},
  volume={103}, number={3}, pages={032430}, year={2021},
  publisher={APS}
}

@article{khan2024brain,
  title={Brain Tumor Diagnosis Using Hybrid Quantum Convolutional Neural Networks},
  author={Khan, Muhammad Al-Zafar and Galib, Abdullah Al Omar and Innan, Nouhaila and Bennai, Mohamed},
  journal={arXiv preprint arXiv:2401.15804}, year={2025}
}

@article{elmaouaki2024quantum,
  title={Quantum Support Vector Machine for Prostate Cancer Detection: A Performance Analysis},
  author={El Maouaki, Walid and Said, Taoufik and Bennai, Mohamed},
  journal={arXiv preprint arXiv:2403.07856}, year={2024}
}

@article{schuld2015introduction,
  author  = {Schuld, Maria and Sinayskiy, Igor and Petruccione, Francesco},
  title   = {An introduction to quantum machine learning},
  journal = {Contemporary Physics},
  volume  = {56},
  number  = {2},
  pages   = {172--185},
  year    = {2015},
  doi     = {10.1080/00107514.2014.964942}
}

@article{holmes2022connecting,
  author  = {Holmes, Zo{\"e} and Sharma, Kunal and Cerezo, M. and Coles, Patrick J.},
  title   = {Connecting Ansatz Expressibility to Gradient Magnitudes and Barren Plateaus},
  journal = {PRX Quantum},
  volume  = {3},
  number  = {1},
  pages   = {010313},
  year    = {2022},
  doi     = {10.1103/PRXQuantum.3.010313}
}

@article{hicks2022evaluation,
  title={On evaluation metrics for medical applications of artificial intelligence},
  author={Hicks, Steven A and Str{\"u}mke, Inga and Thambawita, Vajira and Hammou, Malek and Riegler, Michael A and Halvorsen, P{\aa}l and Parasa, Sravanthi},
  journal={Scientific reports},
  volume={12},
  number={1},
  pages={5979},
  year={2022},
  publisher={Nature Publishing Group UK London}
}

@article{foody2023challenges,
  title={Challenges in the real world use of classification accuracy metrics: From recall and precision to the Matthews correlation coefficient},
  author={Foody, Giles M},
  journal={Plos one},
  volume={18},
  number={10},
  pages={e0291908},
  year={2023},
  publisher={Public Library of Science San Francisco, CA USA}
}

@article{cortes1995support,
  title={Support-vector networks},
  author={Cortes, Corinna and Vapnik, Vladimir},
  journal={Machine learning},
  volume={20},
  number={3},
  pages={273--297},
  year={1995},
  publisher={Springer}
}

@article{breiman2001random,
  title={Random forests},
  author={Breiman, Leo},
  journal={Machine learning},
  volume={45},
  number={1},
  pages={5--32},
  year={2001},
  publisher={Springer}
}

@article{saito2015precision,
  title={The precision-recall plot is more informative than the ROC plot when evaluating binary classifiers on imbalanced datasets},
  author={Saito, Takaya and Rehmsmeier, Marc},
  journal={PloS one},
  volume={10},
  number={3},
  pages={e0118432},
  year={2015},
  publisher={Public Library of Science San Francisco, CA USA}
}

\end{document}